\documentclass[12pt]{iopart}
\usepackage{mathrsfs}
\usepackage{graphicx,color}

\usepackage{iopams}
\def\dag{\dagger}

\newcommand{\ket}[1]{\left| #1 \right\rangle}

\def\bege{\begin{equation}}
\def\ende{\end{equation}}
\def\begen{\begin{eqnarray}}
\def\enden{\end{eqnarray}}

\def\imi{{\bf i}}
\def\gs{{0}}

\begin{document}

\title[]{Sine-square deformation of solvable spin chains and conformal field theories}

\author{Hosho Katsura}

\address{Department of Physics, Gakushuin University, Mejiro, Toshima-ku, Tokyo 171-8588, Japan}
\ead{
\mailto{hosho.katsura@gakushuin.ac.jp},
}
\begin{abstract}
We study solvable spin chains, one-dimensional massless Dirac fermions, and 
conformal field theories (CFTs) with sine-square deformation (SSD), 
in which the Hamiltonian density is modulated by 
the function $f(x)=\sin^2 (\pi x/\ell)$, 
where $x$ is the position and $\ell$ is the length of the system. 
For the XY chain and the transverse field Ising chain at criticality, 
it is shown that the ground state of an open system with SSD 
is identical to that of a uniform chain with periodic boundary conditions.  
The same holds for the massless Dirac fermions with SSD, 
corresponding to the continuum limit of the gapless XY chain. 
For general CFTs, we find that the Hamiltonian of a system with SSD 
has an expression in terms of the generators of the Virasoro algebra. 
This allows us to show that the vacuum state is 
an exact eigenstate of the sine-square deformed Hamiltonian.  
Furthermore, for a restricted class of CFTs associated with affine Lie (Kac-Moody) algebras, 
including $c=1$ Gaussian CFT, 
we prove that the vacuum is an exact ground state of the deformed Hamiltonian. 
This explains why the SSD has succeeded in suppressing boundary effects 
in one-dimensional critical systems, as observed in previous numerical studies. 

\end{abstract}

%\noindent{\it Keywords}: Matrix Product States, Algebraic Bethe Ansatz
%Article preparation, IOP journals
\submitto{\JPA}
\maketitle
%%%%%   Introduction   %%%%%

\section{Introduction}
\label{sec:intro}

The presence of boundaries is usually an obstacle to extracting the bulk properties 
of a system in the thermodynamic limit from analytical or numerical calculations. 
%is also an obstacle to the solvability and/or integrability 
Since the conventional wisdom is that the thermodynamic properties are not affected 
by the boundary conditions imposed,
%much whatever boundary conditions are imposed, 
the periodic boundary conditions (PBC) are often 
used to suppress the finite-size effect caused by the presence of open edges. 
In analytical studies, the boundary conditions other than PBC usually spoil the integrability 
and/or solvability of models. 
However, in practice, in numerical methods such as the density matrix renormalization group (DMRG) method, 
the open boundary conditions (OBC) are more preferable than the PBC, because the calculations under the PBC  
require some additional computational resources. 
%boundary conditions make some significant difference in 

There is a way to reduce the effect of boundaries in the DMRG scheme by turning off 
the interactions smoothly around the edges, which is called smooth boundary conditions~\cite{Vekic1,Vekic2}. 
Similar but modified boundary conditions apply to calculations of transport properties 
such as the conductance of one-dimensional (1D) interacting systems~\cite{Schmitteckert1}. 
More recently, another efficient scheme, which is called the sine-square deformation (SSD), has been proposed~\cite{Gendiar}. 
The SSD rescales the local Hamiltonian density at the center-of-mass position $x$ by the function
\begin{equation}
f_x = \sin^2 \left[ \frac{\pi}{L} \left( x-\frac{1}{2} \right) \right], 
\label{eq:SSD_func}
\end{equation} 
where $L$ is the length of the system and $0 \le x \le L$. 
As seen from $f_{x}=0$ for $x=1/2$ (mod $L$), the SSD breaks the link between the sites $1$ and $L$. 
In various 1D systems including free-fermion chains~\cite{Gendiar}, 
quantum spin chains and ladders~\cite{Hikihara-Nishino}, 
the Hubbard model~\cite{Gendiar2}, and Kondo-lattice model~\cite{Shibata-Hotta}, 
the performance of the SSD has been examined numerically.  
It was found for critical systems that the SSD completely suppresses the boundary effects, i.e., 
the ground-state expectation values of local observables such as the bond strength are translationally invariant.\footnote{
Criticality is not a necessary condition for the success of the SSD. There is a class of gapped models whose ground states are unchanged under the SSD. A typical example is the Affleck-Kennedy-Lieb-Tasaki (AKLT) model  in which the Hamiltonian is a sum of projection operators and the ground state is an eigenstate of each projection~\cite{AKLT_CMP}. 
Note, however, that the unique ground state of the AKLT model with PBC becomes one of the degenerate ground states after the SSD is applied.}
Furthermore, Hikihara and Nishino revealed that the wave-function overlap between 
the ground state of an open system with SSD and that of a uniform system with PBC 
is very close to (almost exactly) unity. 
These observations strongly suggest that the ground state of a uniform 1D system at criticality  
remains almost unchanged even if we introduce the SSD that is spatially modulating by the function $f_x$. 
Motivated by this, a rigorous proof of the correspondence for the spin-$\frac{1}{2}$ XY chain, 
which is equivalent to the system of 1D free fermions, was given by the present author~\cite{Katsura_JPA}. 
A simpler and more physical explanation for the success of the SSD in free-fermion systems was given~\cite{Maruyama-Katsura-Hikihara}, which allows us to extend the idea of the SSD from one dimension to two and higher dimensions.

Although the efficiency of the SSD has been established for free-fermion systems, 
the success of the SSD even in systems with strong interactions such as the XXZ and Hubbard models 
still remains a puzzle and requires further analysis. 
%Therefore, it is natural to ask whether there is a simple mechanism for the success of 
%the SSD even in strongly correlated systems such as the XXZ and the Hubbard models 
%as investigated in previous numerical studies \cite{}. 
In this paper, we reveal the underlying mechanism of the correspondence between 
the systems with PBC and SSD from the perspective of solvable models and conformal field theories (CFTs)
A central role is played by {\it chiral} Hamiltonians introduced below. 
Let us first consider the Hamiltonian for a generic lattice model with nearest-neighbor interaction under PBC: 
\begin{equation}
{\cal H}_0 = \sum^L_{j=1} h_{j,j+1} + \sum^L_{j=1} h_j, 
\end{equation}
where $h_{j,j+1}$ acts only on the sites $j$ and $j+1$ while  $h_j$ on the single site $j$. 
Note that they are periodic in $j$. 
Then, the chiral Hamiltonians are introduced as
\begin{equation}
{\cal H}_\pm = \sum^L_{j=1} e^{\pm \imi \delta j} h_{j,j+1} + \sum^L_{j=1} e^{\pm \imi \delta (j-1/2)} h_j,
\end{equation}
where $\delta =2\pi/L$. The Hamiltonian for the system with SSD is then constructed from 
the original and chiral Hamiltonians as 
\begin{equation}
{\cal H}_{\rm SSD} = \frac{1}{2} {\cal H}_0 - \frac{1}{4} ({\cal H}_+ + {\cal H}_-)
=\sum^{L-1}_{j=1} f_{j+\frac{1}{2}} h_{j,j+1} + \sum^L_{j=1} f_j h_j,
\end{equation}
where $f_x$ is defined in Eq. (\ref{eq:SSD_func}). 
Similarly for continuum field theories, we introduce the uniform and chiral Hamiltonians 
\begin{equation}
{\cal H}_0 = \int^\ell_0 dx\, h(x),~~~~~{\cal H}_\pm =\int^\ell_0 dx\, 
\exp \left( \pm 2\pi \imi \frac{x}{\ell} \right) h(x),
\end{equation}
where $h(x)$ is the Hamiltonian density at $x$ and 
$\ell$ is the circumference of the 1D ring on which the field theory is defined. 
As the Hamiltonian for the continuum theory with SSD, we have
\begin{equation}
{\cal H}_{\rm SSD} = \frac{1}{2} {\cal H}_0 - \frac{1}{4} ({\cal H}_+ + {\cal H}_-)
=\int^\ell_0 dx\, f(x) h(x),
\label{eq:H_SSD_general}
\end{equation}
where
\begin{equation}
\label{eq:func_SSD_cont}
f(x) = \sin^2 \left( \frac{\pi x}{\ell} \right). 
\end{equation}
This Hamiltonian may be regarded as a continuum limit of the lattice system with SSD. 
A common feature of all examples presented in this paper is that the vacuum state of 
the quasi-particles, i.e., the ground state of the original system with PBC, 
is annihilated by the chiral Hamiltonians (${\cal H}_\pm$). This immediately implies that the vacuum state 
is an eigenstate of ${\cal H}_0$ and ${\cal H}_{\pm}$ separately, 
and thus is an exact eigenstate of ${\cal H}_{\rm SSD}$. 
For the isotropic XY and transverse field Ising chains at criticality, 
we can further show that the vacuum state is the unique ground state of ${\cal H}_{\rm SSD}$ 
using the Perron-Frobenius theorem. 
For the model of massless Dirac fermions and a certain class of CFTs, 
${\cal H}_{\rm SSD}$ is found to be positive semidefinite, which means that 
the vacuum is not only an eigenstate but also an exact ground state of ${\cal H}_{\rm SSD}$. 
This may provide a unified explanation for the success of the SSD in both previous numerical 
studies and the analytically solvable lattice models studied in this paper. 

The paper is organized as follows. 
In Sec. 2, we first give some necessary results on the anisotropic XY chain in a field with PBC. 
Then, for the isotropic XY and the transverse field Ising chains at criticality, 
we show that the vacuum state of ${\cal H}_0$ is the unique ground state 
of the Hamiltonian for the system with SSD (${\cal H}_{\rm SSD}$). 
In Sec 3, we consider the model of massless Dirac fermions, which can be regarded as 
an effective field theory of the critical XY chain. We see that the Dirac sea of the uniform system 
still remains an exact ground state after the deformation. 
In Sec 4, we demonstrate that ${\cal H}_{\rm SSD}$ for a generic CFT can be expressed 
as a linear combination of the generators of the Virasoro algebra. 
It turns out that the Hamiltonian involves only the $SL(2, \mathbb{C})$ subalgebra, 
which ensures that the vacuum state of the original CFT is an exact eigenstate. 
For a certain class of CFTs including $c=1$ Gaussian CFT, we show 
that the vacuum state is a ground state of ${\cal H}_{\rm SSD}$. 
We present our conclusions and remarks in Sec. 5. 
The proof of an extension of the Lieb-Schultz-Mattis argument is given in Appendix A. 
In Appendix B, we provide examples of superconformal field theories with SSD.

%%%%%%%%%%%%%%%%%%%%%%%%%%%%%%%%%%%%%%%%%%%%%%%%%%%%%%%%%%%%%%%%%
%%%%%   Definition of the Hamiltonian   %%%%%
\section{Anisotropic XY spin chain with sine-square-deformation}
\label{sec:lattice}
\subsection{Definition of the uniform XY chain}
The anisotropic XY spin chain of length $L$ in a magnetic field is described by the following Hamiltonian:
\begin{equation}
{\cal H}_{0} = -J \sum^{L}_{j=1} [(1+\gamma) S^x_j S^x_{j+1} + (1-\gamma) S^y_j S^y_{j+1}]
-h\sum^L_{j=1} S^z_j, 
\label{eq:Ham_periodic}
\end{equation}
where $S^\alpha_j$ ($\alpha=x,y,z$) are spin-$\frac{1}{2}$ operators on the $j$th site and $\gamma$ is the anisotropy. 
Here, the periodic boundary condition is imposed, i.e., $S^\alpha_{L+1}=S^\alpha_1$. 
For $\gamma=0$, the Hamiltonian reduces to the isotropic XY chain, i.e., the Heisenberg XXZ chain with $\Delta=0$. 
For $\gamma=1$, the Hamiltonian becomes the transverse field Ising chain, which is critical at $h=\pm J$. 
The Hamiltonian for arbitrary values of $\gamma$ and $h/J$ can be solved exactly~\cite{Lieb-Schultz-Mattis, Katsura}. 
In Fig. \ref{fig:FermiPoint}.(a), the phase diagram of the XY chain and the critical lines are shown. 
The XY model has received considerable attention for a long time and various quantities 
such as the correlation functions~\cite{McCoy68, Barouch-McCoy2, Barouch-McCoy4, Its-Izergin-Korepin-Slavnov92}, 
the emptiness formation probability~\cite{Shiroishi-Takahashi-Nishiyama01, Franchini-Abanov05}, 
and the entanglement entropy~\cite{Jin_Korepin_04, Peschel, Its_Jin_Korepin_05, Franchini_Its_Jin_Korepin_07} were calculated exactly. 
%The Hamiltonian with positive exchange $J$ can be obtained from ${\cal H}_{\rm XY}$ 
%by the unitary transformation $U = \prod_{j:{\rm odd}} \exp({\bf i} \pi S^z_j)$: 
%\begin{equation}
%\fl~~~~~~~~~~~~
%U {\cal H}_{\rm XY} U^\dag = J \sum^{L-1}_{j=1} (S^x_j S^x_{j+1} + S^y_j S^y_{j+1})
%+(-1)^L J (S^x_L S^x_1 + S^y_L S^y_1)
%-h\sum^L_{j=1} S^z_j,
%\end{equation}
%where the case of even $L$ corresponds to the PBC while that of odd $L$ corresponds 
%to the antiperiodic boundary condition (APBC) (see footnote 18 of Ref.~\cite{Hikihara-Nishino}). 

%Before going into details, we give a brief synopsis of the known results of 
%the uniform XY chain with PBC, which will be of use later on. 
The Hamiltonian ${\cal H}_{0}$ can be diagonalized through the successive application of 
the Jordan-Wigner and Bogoliubov transformations. 
We first rewrite ${\cal H}_0$ in terms of spinless fermions $c_j$ as
\begin{eqnarray}
%\fl~~~~~~~~~~~
{\cal H}_{0} =& -\frac{J}{2} \sum^{L-1}_{j=1} (c^\dag_j c_{j+1} + c^\dag_{j+1} c_j + \gamma c^\dagger_j c^\dagger_{j+1} + \gamma c_{j+1} c_j)
\nonumber \\
&+\frac{J}{2} \Gamma (c^\dag_L c_{1} + c^\dag_{1} c_L+ \gamma c^\dagger_L c^\dagger_1 + \gamma c_1 c_L)
-h \sum^L_{j=1} \left( c^\dag_j c_j -\frac{1}{2} \right),
\label{eq:Ham_fermion}
\end{eqnarray}
where the Jordan-Wigner fermions are defined through
\begin{eqnarray}
S^+_j = c^\dag_j \prod^{j-1}_{i=1} (1-2c^\dag_i c_i),~~
S^-_j = c_j \prod^{j-1}_{i=1} (1-2c^\dag_i c_i),~~
S^z_j = c^\dag_j c_j -\frac{1}{2}, 
\end{eqnarray}
with $S^\pm_j = S^x_j \pm {\imi} S^y_j$. 
The operator $\Gamma$ in Eq. (\ref{eq:Ham_fermion}) is given by
$
\Gamma := \prod^L_{j=1} (-\sigma^z_j), 
$
which commutes with the Hamiltonian ${\cal H}_{0}$. 
Here, $\sigma^z_j$ is the Pauli matrix and related to $S^z_j$ through $\sigma^z_j =2 S^z_j$. 
Since $\Gamma^2=1$, the eigenstates of ${\cal H}_{0}$ are separated into 
two disconnected sectors with $\Gamma=\pm 1$, in which $+/-$ characterizes 
configurations with an even/odd number of up spins. 
For simplicity, we restrict ourselves here to the case of $\Gamma=+1$, 
in which we have to impose anti-periodic boundary conditions (APBC) on the fermions. 
Note that when $\Gamma=-1$, we have to separately treat the zero-momentum mode, 
which forces the number of fermions in the ground state to be odd. 

To diagonalize the Hamiltonian, we introduce the following linear transformation
\begin{equation}
c_j = \frac{1}{\sqrt L} \sum_k e^{{\imi} kj} (\cos \theta_k d_k + \imi \sin \theta_k d^\dagger_{-k}),
\end{equation}
where the rotation angle $\theta_k$ is defined by 
\begin{equation}
\label{eq:Bog_angle}
\tan (2\theta_k) = -\frac{J \gamma \sin k}{h+J \cos k}. 
\end{equation} 
Note that the following relation holds: $\theta_{-k}=-\theta_k$. 
The new fermion operators satisfy the anti-commutation relations 
$\{ d_k, d^\dagger_{k'} \} = \delta_{k,k'}$, $\{ d_k, d_{k'} \} = \{ d^\dagger_k, d^\dagger_{k'} \} = 0$. 
The APBC implies the quantization of quasimomenta so that $k$ must be in the set
\begin{equation}
{\cal K}:= \left\{ k=\frac{2\pi}{L} \left( n+ \frac{1}{2} \right) \Big|\, n \in \mathbb{Z}\right\},
\end{equation}
where $-\frac{L}{2} \le n \le \frac{L}{2}-1$ for even $L$ and 
$-\frac{L}{2}+\frac{1}{2} \le n \le \frac{L}{2}-\frac{1}{2}$ for odd $L$.  
In terms of the new fermions, the Hamiltonian becomes 
\begin{equation}
{\cal H}_{0} = \sum_{k \in {\cal K}} \epsilon_0 (k) \left( d^\dagger_k d_k - \frac{1}{2} \right),~~
\epsilon_0 (k)= \sqrt{(h+J \cos k)^2+(J \gamma \sin k)^2}.
\end{equation}
The ground state, $\ket{0}$, of ${\cal H}_0$ has no $d$ fermions and hence satisfies $d_k \ket{0}=0$ for all $k$. 

\subsection{XY chain with SSD}
Let us now study the Hamiltonian for the anisotropic XY chain with SSD~\cite{Hikihara-Nishino}. 
To this end, we first introduce the chiral deformation of ${\cal H}_0$ in Eq. (\ref{eq:Ham_fermion}) as follows:
\begin{equation}
{\cal H}_{\pm} = -\frac{J}{2}\sum^L_{j=1} e^{\pm {\imi}\delta j} (c^\dagger_j c_{j+1} +\gamma c^\dagger_j c^\dagger_{j+1} + {\rm H.c.}) -h \sum^L_{j=1} e^{\pm {\imi}\delta (j-1/2)} c^\dagger_j c_j,
\end{equation}
where $\delta = 2\pi/L$. 
Note that ${\cal H}_{\pm}$ are apparently non-Hermitian and their eigenvalues may not be real. 
The Hamiltonian for the system with SSD is then constructed as
\begin{equation}
{\cal H}_{\rm SSD} = \frac{1}{2} {\cal H}_0 -\frac{1}{4}({\cal H}_{+} + {\cal H}_{-}). 
\end{equation}
When $\gamma=0$, the model reduces to the 1D tight-binding model with SSD 
studied in Ref. \cite{Maruyama-Katsura-Hikihara}. 
In the original spin language, the SSD Hamiltonian reads 
\begin{equation}
{\cal H}_{\rm SSD} = -J \sum^{L-1}_{j=1} f_{j+\frac{1}{2}} [(1+\gamma) S^x_j S^x_{j+1} + (1-\gamma) S^y_j S^y_{j+1}]
-h \sum^L_{j=1} f_j S^z_j, 
\label{eq:Ham_SSD}
\end{equation}
where $f_x$ is introduced in Eq. (\ref{eq:SSD_func}). 
As is obvious, this model is defined on the open chain of length $L$, i.e., 
there is no interaction between the end sites ($1$ and $L$). 
This model can also be mapped onto the free spinless fermion system via 
the Jordan-Wigner transformation. However, in contrast to the uniform 
and other known inhomogeneous XY models~\cite{Feldman1, Feldman2, Jeugt1}, 
%the hopping amplitudes and on-site potentials result in position dependent. Therefore, 
the standard Fourier and the Bogoliubov transformations cannot be applied 
and single-particle states cannot be obtained analytically. 
Nevertheless, we can prove that the state $\ket{0}$ is the exact ground state of ${\cal H}_{\rm SSD}$ 
in the following cases: (i) $\gamma=0$ and $h\le J$, and (ii) $\gamma=1$ and $h=J$. 
Note that case (i) has already been studied in Ref.~\cite{Katsura_JPA} and the correspondence was established. 
However, the proof in~\cite{Katsura_JPA} is rather complicated and the second quantized description 
presented below simplifies the proof. Moreover, the new formulation enables us to prove 
the success of the SSD in case (ii) which was not studied in Ref.~\cite{Katsura_JPA}. 
The key to understanding the correspondence in both cases (i) and (ii) 
is the fact that $\ket{0}$ is in the kernel of ${\cal H}_{\pm}$, 
i.e., ${\cal H}_{\pm} \ket{0}=0$. 
This can be easily verified by noting that ${\cal H}_{\pm}$ does not contain terms of the form 
$d^\dagger_k d^\dagger_{k'}$ for cases (i) and (ii). 

\begin{figure}[hbt]
\begin{center}
\includegraphics[width=115mm]{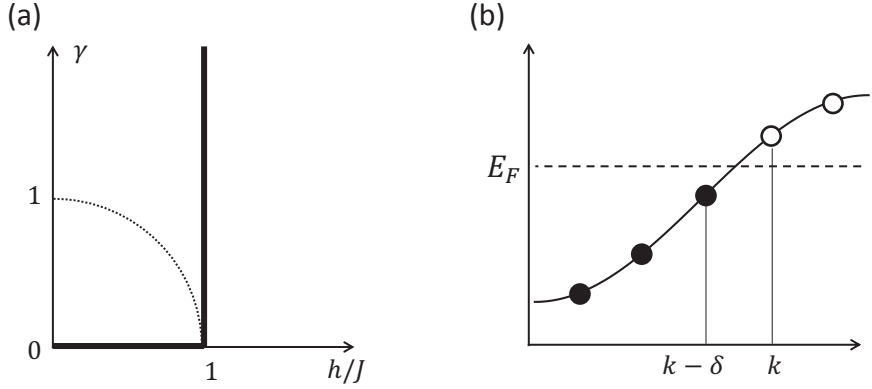}
\caption{
(a) A part of phase diagram of the anisotropic XY model in a magnetic field. 
Note that the model has obvious symmetries $\gamma \to -\gamma$ and $h \to -h$. 
The critical lines are indicated by the bold solid lines. The dotted curve is the Barouch-McCoy circle 
on which the ground state is factorized into a product of single-spin states. 
(b)  A portion of single particle dispersion for case (i). 
The Fermi energy ($E_{\rm F}=0$) is shown by the dashed line.
}
\label{fig:FermiPoint}
\end{center}
\end{figure}
Let us now rewrite the chiral Hamiltonian ${\cal H}_{\pm}$ 
in the basis where ${\cal H}_0$ is diagonal: 
\begin{equation}
\fl~~
{\cal H}_\pm = \frac{1}{2}e^{\mp {\imi} \delta/2} \sum_{k \in {\cal K}}  \left[
{\epsilon}_\pm (k) d^\dagger_{k} d_{k\mp \delta} -{\imi} \eta_\pm (k) d^\dagger_k d^\dagger_{-k \pm \delta} 
+{\imi} \eta_\pm (k) d_{-k} d_{k \mp \delta} - {\epsilon}_\pm(k) d_{-k} d^\dagger_{-k \pm \delta},
\right]
\end{equation}
where $\epsilon_\pm (k)$ and $\eta_\pm (k)$ are given by
\begin{eqnarray}
%{\epsilon}_\pm (k) &= -[h+J \cos(k \mp \delta/2) ] (u_k u_{k\mp \delta} - v_k v_{k \mp \delta})
%\nonumber \\
%&+ J \gamma \sin (k \mp \delta/2) (u_k v_{k\mp \delta} - v_k u_{k \mp \delta}),
%\\
%{\cal \eta}_\pm (k) &= [h+J \cos(k \mp \delta/2)] (u_k v_{k\mp \delta} + v_k u_{k \mp \delta})
%\nonumber \\
%&+J \gamma \sin (k \mp \delta/2) (u_k u_{k\mp \delta} - v_k v_{k \mp \delta}).
\fl~~~~~
\epsilon_\pm (k) = -[h+J \cos(k \mp \delta/2)] \cos (\theta_k + \theta_{k \mp \delta}) 
+ J \gamma \sin (k \mp \delta/2) \sin (\theta_k + \theta_{k \mp \delta})
\\
\fl~~~~~
\eta_\pm (k) = +[h+J \cos(k \mp \delta/2) ] \sin (\theta_k + \theta_{k \mp \delta})
+ J \gamma \sin (k \mp \delta/2) \cos (\theta_k + \theta_{k \mp \delta})
\end{eqnarray}
Using the relations
\begin{equation}
\label{eq:cos_sin}
\cos (2\theta_k) = -\frac{h+J \cos k}{\epsilon_0 (k)},~~~~~
\sin (2\theta_k) = \frac{J \gamma \sin k}{\epsilon_0 (k)},
\end{equation}
one can further rewrite $\eta_\pm (k)$:
\begin{equation}
\label{eq:eta_pm}
\eta_\pm (k) = - \epsilon_0 (k \mp \delta/2) \sin (\theta_k -2 \theta_{k \mp \delta/2} + \theta_{k \mp \delta}). 
\end{equation}
To see that $\ket{0}$ is annihilated by ${\cal H}_\pm$ for cases (i) and (ii), 
we have only to show that $\eta_\pm (k)=0$ for all $k$. 
We shall discuss these two cases separately. 
For (i) where $\gamma=0$ and $h\le J$, the model reduces to the isotropic XY chain in a magnetic field, 
which has been studied in Ref. \cite{Katsura_JPA}. 
In this case, we can explicitly solve Eq. (\ref{eq:cos_sin}) and obtain
\begin{equation}
\theta_k = \left\{ 
\begin{array}{c}
0~~~~~~~~h+J\cos k < 0 \\
\frac{\pi}{2}~~~~~~~~h+J\cos k >0.
\end{array}
\right.
\end{equation}
Therefore, $\eta_\pm (k)=0$ except for $(h+J \cos k) [h + J\cos (k \mp \delta)]<0$, 
and the state $\ket{0}$ is at least an approximate eigenstate of ${\cal H}_\pm$. 
As discussed in Refs. \cite{Katsura_JPA, Maruyama-Katsura-Hikihara},  
one can make $\ket{0}$ an exact eigenstate by fine-tuning the magnetic field 
(chemical potential for spinless fermions) so that $\epsilon_0 (k \mp \delta/2)=0$ 
when the Fermi point is in between $k$ and $k \mp \delta$ (see Fig. \ref{fig:FermiPoint} (b)). 
This condition implies that the chiral Hamiltonian, a nearest-neighbor tight-binding model 
in reciprocal space, does not contain terms with momentum transfer across the Fermi points 
(for details, see Ref.~\cite{Maruyama-Katsura-Hikihara}). 
The corresponding magnetic field reads $h = -J \cos (N\pi/L)$, with $N$ being the number of up spins. 
%, where $N$ is the number of up spins. 
For (ii) where $\gamma=1$ and $h=J$, we can again explicitly solve Eq. (\ref{eq:Bog_angle}) and obtain
\begin{equation}
\theta_k = -\frac{k}{4}.
\end{equation}
This yields $\theta_k -2 \theta_{k \mp \delta/2}+\theta_{k \mp \delta}=0$ and hence $\eta_\pm (k)=0$ for all $k$. 
Therefore, when (i) $\gamma=0$ and $h/J=-\cos(N\pi/L)$, or (ii) $\gamma=1$ and $h/J=1$, 
the state $\ket{0}$ is annihilated by ${\cal H}_\pm$, which implies that $\ket{0}$ is an exact eigenstate of 
${\cal H}_{\rm SSD}$ with the eigenvalue $E_0/2$, where $E_0$ is the ground-state energy of ${\cal H}_0$. 
%\textbf{[H.K.: When $\gamma \to \infty$, $\theta_k=-{\rm sgn}(k) \pi/4$ and $\eta_\pm (k)=0$ is satisfied for all $k$. 
%A similar simplification occurs when $h/J \to \infty$. Those cases include gapped cases. Is it contradicting?]} 

Several comments are in order. 
It is natural to ask whether we can find other points in the parameter space 
at which $\ket{0}$ is annihilated by ${\cal H}_\pm$. 
One can actually solve Eq. (\ref{eq:cos_sin}) explicitly in the following two limits: 
(iii) $\gamma \to \infty$, and (iv) $h/J \to \infty$, 
and can find that the condition $\eta_\pm (k) =0$ is satisfied. 
However, the Hamiltonian for the limit (iii) (with appropriate scaling) is achieved from the isotropic XY chain 
by the unitary transformation which rotates every other spin by angle $\pi$ about the $x$-axis, 
i.e., $(S^x_j, S^y_j, S^z_j) \to (S^x_j, -S^y_j, -S^z_j)$ for every other $j$. 
Therefore, the argument for case (i) equally applies to (iii). 
In the limit (iv), the ground state of the uniform Hamiltonian is a simple product state and 
is insensitive to the spatial modulation induced by the SSD. 
Another question is what is the ground state and the nature of the low-lying states of 
the SSD Hamiltonian (${\cal H}_{\rm SSD}$) for cases (i) and (ii). 
As we will see in the next subsection, the exact eigenstate $\ket{0}$ turns out to be 
the unique ground state of ${\cal H}_{\rm SSD}$ in the two cases (i) and (ii). 
Furthermore, for case (i), we can show that the lowest excitation energy of ${\cal H}_{\rm SSD}$ 
is $O(1/L)$ using an extension of the Lieb-Schultz-Mattis argument~\cite{Lieb-Schultz-Mattis, Affleck-Lieb}, 
which suggests the existence of gapless excitation in the limit of $L \to \infty$. 
The proof is given in Appendix A.

\subsection{Uniqueness of the ground state}
%So far, we have shown that the vacuum for the quasi-particles, $\ket{0}$, is an exact eigenstate of ${\cal H}_{\rm SSD}$ for cases (i) and (ii). 
In this subsection, we shall prove that $\ket{0}$ is the unique ground state of ${\cal H}_{\rm SSD}$. 
Since the statement has been established for case (i) 
%uniqueness of the ground state and the correspondence for case (i)
in Ref. \cite{Katsura_JPA}, here we restrict our attention to case (ii). 
The Hamiltonian for case (ii) is written as
\begin{equation}
\label{eq:SSD_TFI}
{\cal H}_{\rm SSD} = -2J \sum^{L-1}_{j=1} f_{j+\frac{1}{2}} S^x_j S^x_{j+1} 
-J \sum^L_{j=1} f_j S^z_j, 
\end{equation}
where $J>0$ is assumed. 
To make use of the Perron-Frobenius theorem, we have to use the basis in which $S^x_j$ is diagonalized:
\begin{equation}
\ket{\rightarrow}_j := (\ket{\uparrow}_j+\ket{\downarrow}_j)/\sqrt{2},~~~~
\ket{\leftarrow}_j := (\ket{\uparrow}_j-\ket{\downarrow}_j)/\sqrt{2}.
\end{equation}
In this basis, from $f_j >0$ for all $j=1, 2,..., L$, 
all of the off-diagonal elements of ${\cal H}_{\rm SSD}$ are nonpositive and satisfy the connectivity condition. 
Therefore, the Perron-Frobenius theorem tells us that the ground state is nondegenerate 
and can be written in the form
\begin{equation}
\label{eq:positive_exp}
\ket{\Psi_{\gs}} = \sum_{\alpha_1, ..., \alpha_L=\rightarrow, \leftarrow} c_{\alpha_1, ..., \alpha_L} 
\ket{\alpha_1}_1 \otimes ... \otimes \ket{\alpha_L}_L,
\end{equation}
where the coefficients $c_{\alpha_1, ..., \alpha_L}$ are strictly positive for any $(\alpha_1, ..., \alpha_L)$.  
Obviously, the same argument holds for the uniform transverse field Ising model at criticality 
and hence the state $\ket{0}$ has the same property. 
A nice corollary of this result is that the eigenvalue of the operator $\Gamma$  
is $\pm 1$ in the ground state of a chain of even/odd length $L$ for both uniform and SSD chains. 

Let us now focus on the case of even $L$. In this case, we have shown in the previous subsection 
that $\ket{0}$ is an exact eigenstate of ${\cal H}_{\rm SSD}$. Since there can be no other state having only positive coefficient that is orthogonal to $\ket{\Psi_{\gs}}$, $\ket{0}$ is identical to $\ket{\Psi_{\gs}}$ apart from an overall factor. 
This proves that $\ket{0}$ is the unique ground state of ${\cal H}_{\rm SSD}$ (Eq. (\ref{eq:SSD_TFI})).

\section{Dirac Hamiltonian with SSD}
\label{sec:Dirac}
\begin{figure}[bht]
\begin{center}
\includegraphics[width=115mm]{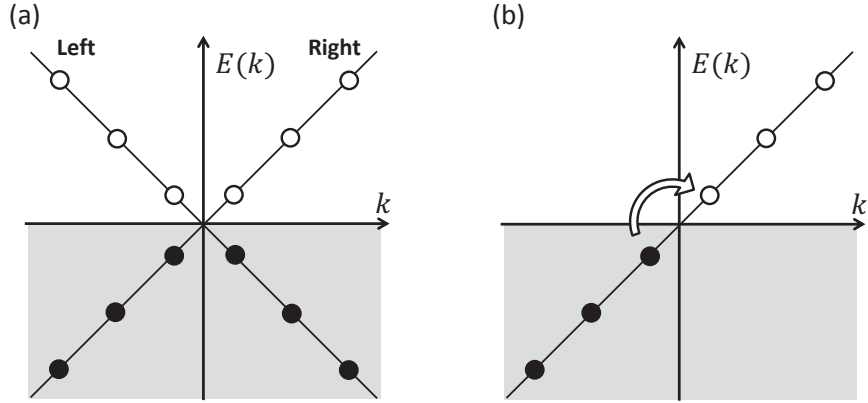}
\caption{
(a) Dispersion relation of the Dirac fermions. All negative energy levels are filled in the Dirac sea. 
The momentum $k$ takes discrete values $2\pi n/\ell$ ($n \in \mathbb{Z}+\frac{1}{2}$). 
(b) Action of $\psi^\dagger_{{\rm R}, 1/2} \psi_{{\rm R}, -1/2}$ on the Dirac sea, which is 
excluded in ${\cal H}_+$. 
}
\label{fig:DiracSea}
\end{center}
\end{figure}

In this section, we shall consider the massless Dirac fermion system with SSD. 
The original periodic system can be regarded as an effective field theory of 
the XY chain or an equivalent spinless-fermion model. 
The Hamiltonian of the massless Dirac fermions consists of the Left(L)- and Right(R)-moving ones:
\begin{equation}
{\cal H}_0 = 
%{\cal H}_{\rm L} + {\cal H}_{\rm R},~~
\imi \frac{v_{\rm F}}{2\pi} \int^\ell_0 dx\, \left[ :\psi^\dagger_{\rm L}(x) \frac{d}{dx} \psi_{\rm L}(x):
-:\psi^\dagger_{\rm R}(x) \frac{d}{dx} \psi_{\rm R}(x): \right],
\end{equation}
where $v_{\rm F}$ is the Fermi velocity, $\ell$ is the circumference of the ring, 
and the symbol $:~:$ denotes the standard normal ordering 
(see Ref.~\cite{Ludwig_lectures} for details). 
To ensure the uniqueness of the ground state, we impose the anti-periodic boundary condition on fermions, 
i.e., $\psi_{\rm L/R}(x+\ell)=-\psi_{\rm L/R}(x)$. 
Then, the fermion fields are expanded as
\begin{equation}
\psi_{\rm L} (x) = \sqrt{\frac{2\pi}{\ell}} \sum_{n \in \mathbb{Z}+\frac{1}{2}} e^{{\imi} \delta n x} \psi_{{\rm L},n},~~
\psi_{\rm R} (x) = \sqrt{\frac{2\pi}{\ell}} \sum_{n \in \mathbb{Z}+\frac{1}{2}} e^{{\imi} \delta n x} \psi_{{\rm R},n},
\end{equation}
where $\delta=2\pi/\ell$ 
and the operators $\psi_{{\rm L},n}$ and $\psi_{{\rm R},n}$ obey the anti-commutation relations 
$\{ \psi_{{\rm L/R},m}, \psi^\dagger_{{\rm L/R},n} \} = \delta_{m,n}$. 
The Hamiltonian in momentum space is written as
\begin{equation}
{\cal H}_0 = \frac{2\pi}{\ell}v_{\rm F} \sum_{n \in \mathbb{Z}+\frac{1}{2}}  
(-n:\psi^\dagger_{{\rm L},n} \psi_{{\rm L},n}: + n :\psi^\dagger_{{\rm R},n} \psi_{{\rm R},n}: ). 
\end{equation}
By carefully examining the normal ordering, one can see that the zero-energy ground state of ${\cal H}_0$ 
is the Dirac sea where all negative energy levels are occupied (see Fig. \ref{fig:DiracSea}). 

Let us now consider the system with SSD. 
We will see that the Dirac sea, ${\ket {\rm DS}}$, is an exact ground state of the Hamiltonian with SSD. 
We introduce the chiral deformation of the original Hamiltonian 
in a form reminiscent of the lattice model one:
\begin{eqnarray}
{\cal H}_{\pm} &=& {\imi} \frac{v_{\rm F}}{2\pi} \int^\ell_0 dx\, 
e^{\pm {\imi}\delta x} \left[ :\psi^\dagger_{\rm L}(x) \frac{d}{dx} \psi_{\rm L}(x):
-:\psi^\dagger_{\rm R}(x) \frac{d}{dx} \psi_{\rm R}(x): \right]
\nonumber \\
&-& \mu_\pm \frac{1}{2\pi} \int^\ell_0 dx\, e^{\pm {\imi}\delta x}
\left[ :\psi^\dagger_{\rm L}(x) \psi_{\rm L}(x): - :\psi^\dagger_{\rm R}(x) \psi_{\rm R}(x): \right],
\end{eqnarray}
where the chemical potential $\mu_\pm=\pm \pi v_{\rm F}/\ell$ is introduced in such a way that 
the relation $({\cal H}_+)^\dagger = {\cal H}_-$ holds. 
Substituting the mode expansions of the fermions, we arrive at
\begin{equation}
{\cal H}_{\pm} = \frac{2\pi}{\ell} v_{\rm F} \sum_{n \in \mathbb{Z}+\frac{1}{2}} 
\left[ -\left( n\pm \frac{1}{2} \right) \psi^\dagger_{{\rm L}, n \pm 1} \psi_{{\rm L}, n}
+ \left( n \pm \frac{1}{2} \right) \psi^\dagger_{{\rm R}, n \pm 1} \psi_{{\rm R}, n} \right]. 
\end{equation}
Similar to the chiral deformation of the lattice fermion systems~\cite{Maruyama-Katsura-Hikihara}, 
${\cal H}_{\pm}$ is expressed in terms of the sum of nearest-neighbor hoppings in momentum space. 
More importantly, ${\cal H}_{\pm}$ does not include the momentum transfer across the Fermi point (see Fig. \ref{fig:DiracSea} (b)).  
In addition, in the Dirac sea, momentum transfers occurring below the Fermi energy are prohibited 
by the Pauli exclusion principle. 
Therefore, the state $\ket{\rm DS}$ is annihilated by ${\cal H}_{\pm}$ and hence the zero-energy 
eigenstate of it. 

We are now in a position to consider the Hamiltonian for the system with SSD. 
The Hamiltonian is constructed from the original and chiral ones as
\begin{eqnarray}
{\cal H}_{\rm SSD} &=& \frac{1}{2} {\cal H}_0 - \frac{1}{4} ({\cal H}_+ + {\cal H}_-) 
\nonumber \\
&=& {\imi} \frac{v_{\rm F}}{2\pi} \int^\ell_0 dx\, f(x) \left[ :\psi^\dagger_{\rm L}(x) \frac{d}{dx} \psi_{\rm L}(x):
-:\psi^\dagger_{\rm R}(x) \frac{d}{dx} \psi_{\rm R}(x): \right]
\nonumber \\
&+& \frac{v_{\rm F}}{4 \ell} \int^\ell_0 dx\, \cos \left( \frac{2\pi}{\ell}x \right)
\left[ :\psi^\dagger_{\rm L}(x) \psi_{\rm L}(x): - :\psi^\dagger_{\rm R}(x) \psi_{\rm R}(x): \right]
\label{eq:H_SSD_Dirac}
\end{eqnarray}
where $f(x)=\sin^2 (\pi x/\ell)$ is the scaling function. 
Because $\ket{\rm DS}$ is a simultaneous eigenstate of ${\cal H}_0$ and ${\cal H}_{\pm}$ 
with zero eigenenergy, it is also a zero-energy eigenstate of ${\cal H}_{\rm SSD}$. 
The second term in Eq. (\ref{eq:H_SSD_general}) becomes negligible for large circumference $\ell$. 
%Nevertheless, this term is essential for ensuring Hermiticity of ${\cal H}_{\rm SSD}$. 
%One might think that the Hamiltonian in Eq. (\ref{eq:H_SSD_Dirac}) shows a deviation 
%from the general definition of ${\cal H}_{\rm SSD}$ in Eq. (\ref{eq:H_SSD_general}). 
%This conflict can be resolved by adding the following term to ${\cal H}_0$:
%\begin{equation}
%-\frac{v_{\rm F}}{2\ell} \int^\ell_0 dx \, 
%\left[ :\psi^\dagger_{\rm L}(x) \psi_{\rm L}(x): - :\psi^\dagger_{\rm R}(x) \psi_{\rm R}(x): \right].
%\end{equation}
%Note that this additional term only shifts the energy levels of ${\cal H}_0$ by a constant 
%and does not affect the conclusion that the Dirac sea is an eigenstate of the deformed Hamiltonian. 
Since $\ket{\rm DS}$ is a simultaneous eigenstate of ${\cal H}_0$ and ${\cal H}_{\pm}$ 
with zero eigenenergy, it is also a zero-energy eigenstate of ${\cal H}_{\rm SSD}$. 
One can further show that the Dirac sea is a ground state of ${\cal H}_{\rm SSD}$. 
To see this, we rewrite the Hamiltonian in terms of positive-semidefinite operators as
\begin{eqnarray}
\fl~~~~~
{\cal H}_{\rm SSD} = \frac{\pi v_{\rm F}}{2\ell} \sum_{n \in \mathbb{Z}+\frac{1}{2}, n>0}
\left( n+\frac{1}{2} \right) (A_{{\rm L},n} A^\dagger_{{\rm L},n} + B^\dagger_{{\rm L},n} B_{{\rm L},n} 
+ A^\dagger_{{\rm R},n} A_{{\rm R},n} + B_{{\rm R},n} B^\dagger_{{\rm R},n} ),
%[ (\psi_{{\rm L},n}-\psi_{{\rm L},n+1}) (\psi^\dagger_{{\rm L},n}-\psi^\dagger_{{\rm L},n+1})  
%+ (\psi^\dagger_{{\rm L},-n}-\psi^\dagger_{{\rm L},-n-1}) (\psi_{{\rm L},-n}-\psi_{{\rm L},-n-1}) 
%\nonumber \\
%\fl
%&+ (\psi^\dagger_{{\rm R},n}-\psi^\dagger_{{\rm R},n+1}) (\psi_{{\rm R},n}-\psi_{{\rm R},n+1})
%+ (\psi_{{\rm L},-n}-\psi_{{\rm L},-n-1}) (\psi^\dagger_{{\rm L},-n}-\psi^\dagger_{{\rm L},-n-1})]. 
\end{eqnarray}
where
\begin{equation}
A_{{\rm L/R},n} = \psi_{{\rm L/R},n} - \psi_{{\rm L/R},n+1},~~~~~
B_{{\rm L/R},n} = \psi_{{\rm L/R},-n}-\psi_{{\rm L/R},-n-1}.
\end{equation}
From this decomposition, one can easily see that $\langle \Psi| {\cal H}_{\rm SSD} \ket{\Psi} \ge 0$ 
for all state $\ket{\Psi}$, 
yielding that all the eigenvalues of ${\cal H}_{\rm SSD}$ are nonnegative. 
This proves that the ground state of the homogeneous Dirac Hamiltonian, $\ket{\rm DS}$, is 
also an exact ground state of the Hamiltonian with SSD. 
A similar strategy will be used in the next section to show the positive semidefiniteness of the SSD Hamiltonian 
in a wide class of (1+1)-dimensional critical systems. 
We finally remark that the SSD also works in a system comprised of purely left- or right- moving Dirac fermions, 
called {\it chiral} Dirac fermions,  which is relevant to understanding the edge state in the quantum Hall effect
~\cite{Prange_QHE}. 

\section{CFT with SSD}
\label{sec:CFT}

So far, we have seen that the SSD leaves the ground state unchanged in the anisotropic XY chain at criticality 
as well as in the Dirac fermion system, which is a representative of $c=1$ CFT. 
In this section, we shall consider more general CFTs with SSD. 
In fact, we will find that the Hamiltonian for a generic CFT with SSD can be expressed in terms of 
the generators of the Virasoro algebra, which accounts for the success of the SSD in a wide class 
of the (1+1)-dimensional critical systems. 
\begin{figure}[bht]
\begin{center}
\includegraphics[width=100mm]{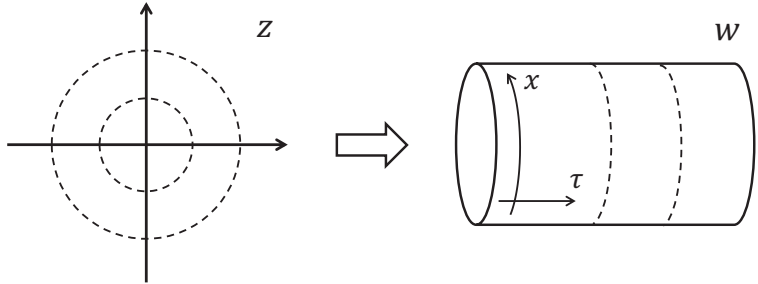}
\caption{
Conformal mapping from the complex plane to the cylinder.
}
\label{fig:cylinder}
\end{center}
\end{figure}

Let us consider CFT on an infinitely long cylinder of circumference $\ell$. 
We denote the coordinate along the cylinder by $\tau$ while that along the circumference by $x$. 
As shown in Fig. \ref{fig:cylinder}, the space coordinate $x$ is compactified as $0 \le x \le \ell $. 
The Hamiltonian on the cylinder is defined as
\begin{equation}
\label{eq:CFT_ham}
{\cal H}_0 = \int^\ell_0 \frac{d x}{2\pi} (T_{\rm cyl}(w) + {\overline T}_{\rm cyl} ({\overline w})),
\end{equation}
where $w=\tau + {\bf i} x$ and $T_{\rm cyl}(w)$ is the energy-momentum tensor on the cylinder. 
Using the transformation property of the energy-momentum tensor under the conformal mapping, 
\begin{equation}
w=\tau + {\bf i} x = \frac{\ell}{2\pi} \log z, 
\end{equation}
one can obtain the relation between $T_{\rm cyl}(w)$ and the energy-momentum tensor 
on the entire complex plane ($z$)~\cite{YellowBook, Mussardo}
\begin{equation}
T_{\rm cyl} (w) = \left(\frac{2 \pi}{\ell} \right)^2 \left[ T(z) z^2 - \frac{c}{24} \right], 
\end{equation}
where $c$ denotes the central charge and 
the constant term comes from the Schwarzian derivative. 
As is well known, $T(z)$ can be expanded by the Virasoro generators as
\begin{equation}
\label{eq:stress}
T(z) = \sum_{n \in \mathbb{Z}} z^{-n-2} L_n
\end{equation}
and similarly for ${\overline T}({\overline z})$.  
The Virasoro generators satisfy the following commutation relation:
\begin{equation}
[L_m, L_n] = (m-n) L_{m+n} + \frac{c}{12} (m^3-m) \delta_{m+n,0}. 
\end{equation}
The Hilbert space of the CFT is constructed by the actions of 
the generators on the vacuum state (and the other primary states). 
For the vacuum state, we demand $L_m \ket{0} = 0$ when $m \ge -1$.  
In other words, the vacuum is $SL(2,\mathbb{C})$ invariant, i.e.,  
the operators $\{L_0, L_1, L_{-1} \}$ annihilating $\ket{0}$ constitute the $SL(2,\mathbb{C})$ subgroup of the conformal group. 
Note that the true physical vacuum is the tensor product $\ket{\rm vac}=\ket{0} \otimes \ket{\overline 0}$. 
Substituting Eq. (\ref{eq:stress}) into (\ref{eq:CFT_ham}), we have
\begin{equation}
{\cal H}_0 = \frac{2\pi}{\ell} (L_0 +{\overline L}_0) -\frac{\pi c}{6 \ell}. 
\end{equation}
As is obvious, the vacuum state $\ket{\rm vac}$ is the ground state of ${\cal H}_0$ 
with the finite-size correction to the ground state energy $E_0=-\frac{\pi c}{6 \ell}$~\cite{Bloete86, Affleck86_PRL}.

Let us next derive the relation between the chiral Hamiltonian and the Virasoro generators. 
As we have seen in the previous sections, the chiral Hamiltonian is nothing but 
the Fourier component of the energy density with the wavenumber $\pm 2\pi/\ell$. 
Analogous to Eq. (\ref{eq:CFT_ham}), we define the chiral Hamiltonian as
\begin{equation}
{\cal H}_{\pm} = \int^\ell_0 \frac{d x}{2\pi} (e^{\pm \delta w} T_{\rm cyl}(w) + e^{\mp \delta {\overline w}}\, {\overline T}_{\rm cyl} ({\overline w})),
\end{equation}
where $\delta = 2\pi/\ell$. 
Again using the mode expansion of $T(z)$ (and similar one for ${\overline T}({\overline z})$), we arrive at
\begin{equation}
{\cal H}_{\pm} = \frac{2\pi}{\ell} (L_{\pm 1} +{\overline L}_{\mp 1}).
\end{equation}
Therefore, the Hamiltonian for a generic CFT with SSD can be expressed in terms of 
the Virasoro generators as
\begin{eqnarray}
\label{eq:CFT_SSD_ham}
{\cal H}_{\rm SSD} 
= \frac{1}{2} {\cal H}_0 - \frac{1}{4}({\cal H}_{+} + {\cal H}_{-}) 
= {\cal H}_{\rm L} + {\cal H}_{\rm R}- \frac{\pi c}{12 \ell}
%&= \frac{\pi}{\ell} (L_0 + {\overline L}_0) -\frac{\pi}{2 \ell} (L_{1} + L_{-1} +{\overline L}_{1} + {\overline L}_{-1}) - \frac{\pi c}{12 \ell}.
\end{eqnarray}
with 
\begin{equation}
\label{eq:CFT_L_ham}
{\cal H}_{\rm L} = \frac{\pi}{\ell} \left(L_0 -\frac{L_1+L_{-1}}{2} \right),~~
{\cal H}_{\rm R} = \frac{\pi}{\ell} \left({\overline L}_0 -\frac{{\overline L}_1+{\overline L}_{-1}}{2} \right).
\end{equation}
This is the central result of our paper. 
From the $SL(2,\mathbb{C})$ invariance of the vacuum state, i.e., $L_0\ket{0}=L_{\pm 1}\ket{0}=0$, 
it immediately follows that 
the vacuum state of the original periodic system is an exact eigenstate of ${\cal H}_{\rm SSD}$
\begin{equation}
{\cal H}_{\rm SSD} \ket{\rm vac} = \frac{E_0}{2} \ket{\rm vac}~~{\rm with}~~E_0=-\frac{\pi c}{6 \ell}.
\end{equation}
%Here we have assumed that the SSD does not lead to a specific conformal boundary state, 
%which implies that $\ket{\rm vac}$ is a state in the Hilbert space. 
The relation above clearly shows that $\ket{\rm vac}$ has the eigenenergy obeying
the same scaling relation as that for the ground state of ${\cal H}_0$.

For a class of CFTs associated with affine Lie algebras (or Kac-Moody algebras), 
one can further show that the vacuum state $\ket{\rm vac}$ is an exact ground state of ${\cal H}_{\rm SSD}$. 
To see this, let us first consider the $c=1$ CFT as the simplest example. 
%To show this,
We recall the bosonization of the Virasoro algebra~\cite{YellowBook}. 
In terms of the generators of the Heisenberg algebra with the relation 
$
[a_n, a_m]=n \delta_{n+m,0}, 
$
the Virasoro generators associated with $c=1$ CFT can be expressed as
\begin{eqnarray}
L_n = \frac{1}{2} \sum_{n \in \mathbb{Z}} : a_m a_{n-m}:,
\end{eqnarray}
where the symbol $:~:$ denotes the normal ordering. 
One can also obtain a similar expression for ${\overline L}_n$ using 
the generators ${\overline a}_n$ with the relations $[{\overline a}_n, {\overline a}_m]=n \delta_{n+m,0}$ 
and $[a_n, {\overline a}_m]=0$. 
The Fock space is generated from the charged vacuum $\ket{\alpha}$ by the actions of the creation operators $a_n$ ($n<0$). 
Note that $a_n \ket{\alpha}=0$ ($n > 0$) and $a_0 \ket{\alpha}=\alpha\ket{\alpha}$. 
%The parameter $\rho$ and the central charge $c$ are related through $c=1-12\rho^2$. 
We shall now show that the eigenvalue of the SSD Hamiltonian ${\cal H}_{\rm SSD}$ 
in Eq. (\ref{eq:CFT_SSD_ham}) is bounded below by $E_0/2$. 
To see this, we first rewrite ${\cal H}_L$ in Eq. (\ref{eq:CFT_L_ham}) as
\begin{eqnarray}
\label{eq:psd_ham}
{\cal H}_{\rm L} &=& \frac{\pi}{2\ell} \sum_{n \ge 0} (a_{-n}-a_{-n-1}) (a_n-a_{n+1})
\nonumber \\
                                     &=& \frac{\pi}{2\ell} \sum_{n \ge 0} (a^\dagger_n-a^\dagger_{n+1}) (a_n-a_{n+1}),
\end{eqnarray}
where we have used the fact that $\alpha$ is a real number. 
We can also obtain a similar expression for ${\cal H}_{\rm R}$ by replacing $a_n$ with ${\overline a}_n$. 
Since ${\cal H}_{\rm L}$ and ${\cal H}_{\rm R}$ are expressed in terms of the sum of positive semidefinite operators, 
all the eigenvalue of ${\cal H}_{\rm L} + {\cal H}_{\rm R}$ are nonnegative. 
Another crucial point is that ${\cal H}_{\rm L}+{\cal H}_{\rm R}$ does not contain terms of the form 
$a_m a_n$ with negative $m$ and $n$, which is similar to what we have seen in the anisotropic XY chain, 
in which $\eta_\pm (k)$ in Eq. (\ref{eq:eta_pm}) is zero when the SSD leaves the ground state unchanged. 
Therefore, the eigenvalue of ${\cal H}_{\rm SSD}$ is bounded below by $E_0/2$ 
and the vacuum state $\ket{\rm vac}$ saturating this bound is a ground state of ${\cal H}_{\rm SSD}$. 
This proves that the SSD does not alter the ground state (vacuum) of the original $c=1$ CFT. 
Note, however, that the uniqueness of the ground state of ${\cal H}_{\rm SSD}$ in Eq. (\ref{eq:psd_ham}) 
is an open issue. 

It is well known that the physical realization of the $c=1$ CFT is the Gaussian (free-boson) field theory, 
which is the bosonized theory of the Dirac fermions studied in the previous section.  
Therefore, it is natural to expect that the local Hamiltonian of this CFT with SSD 
is modified from the Gaussian one according to the function $f(x)$ introduced in Eq. (\ref{eq:func_SSD_cont}).  
In order to expose ${\cal H}_{\rm SSD}$ in real space, 
we introduce the mode expansions of the Bose fields~\cite{Eggert-Affleck, Bortz-Eggert, Furukawa-Sato-Furusaki}:
\begin{eqnarray}
\phi_{\rm L} (x) &=& \phi_{{\rm L},0} + Q_{\rm L} \frac{x}{\ell} 
+\frac{\bf i}{\sqrt{4\pi}} \sum^\infty_{n=1} \frac{1}{n} (e^{-{\bf i}\delta n x}a_n - e^{{\bf i}\delta n x} a_{-n} ),  
\\
\phi_{\rm R} (x) &=& \phi_{{\rm R},0} + Q_{\rm R} \frac{x}{\ell}
+\frac{\bf i}{\sqrt{4 \pi}} \sum^\infty_{n=1} \frac{1}{n} (e^{{\bf i}\delta n x} {\overline a}_n - e^{-{\bf i}\delta n x} {\overline a}_{-n} ),
\end{eqnarray}
where $\delta=2\pi/\ell$, and the operators $\phi_{{\rm L/R},0}$ and $Q_{\rm L/R}$ satisfy 
$[\phi_{{\rm L},0}, \phi_{{\rm R},0}]=-{\bf i}/2$ and $[Q_{\rm L/R}, \phi_{{\rm L/R},0}]=\mp {\bf i}/2$. 
Note the bosonic fields obey the commutation relations $[\phi_{\rm L/R}(x), \phi_{\rm L/R}(y)] = \mp {\bf i} {\rm sign}(x-y)/4$, 
and $[ \phi_{\rm L}(x), \phi_{\rm R}(y) ] = -{\bf i}/2$. 
%\textbf{[H.K.: Is it correct? Anyway, it's not so important...]}. 
One can easily see that the Virasoro generators are expressed in terms of the bosonic fields as
\begin{eqnarray}
\fl~~~~~~~~~~
\frac{2\pi}{\ell} L_0 = \int^\ell_0 dx\, :\left( \frac{d \phi_{\rm L}}{d x} \right)^2:,~~
\frac{2\pi}{\ell} {\overline L}_0 = \int^\ell_0 dx\, :\left( \frac{d \phi_{\rm R}}{d x} \right)^2:,
\nonumber \\
\fl~~~~~~~~~~
\frac{2\pi}{\ell} L_{\pm 1} = \int^\ell_0 dx\, e^{\pm {\bf i} \delta x} :\left( \frac{d \phi_{\rm L}}{d x} \right)^2:,~~
\frac{2\pi}{\ell} {\overline L}_{\mp 1} = \int^\ell_0 dx\, e^{\pm {\bf i} \delta x} :\left( \frac{d \phi_{\rm R}}{d x} \right)^2:,
\end{eqnarray}
where we have used the relations $Q_{\rm L}=a_0/\sqrt{\pi}$ and $Q_{\rm R}={\overline a}_0/\sqrt{\pi}$. 
Substituting the above expressions into Eq. (\ref{eq:CFT_SSD_ham}), 
the SSD Hamiltonian in real space reads
\begin{equation}
\label{eq:gaussian_SSD}
{\cal H}_{\rm SSD} = \int^\ell_0 dx\, f(x) h(x) -\frac{\pi}{12 \ell}
\end{equation}
with 
\begin{equation}
f(x) = \sin^2 \left(\frac{\pi x}{\ell} \right),~~ h(x)= :\left(\frac{d \phi_{\rm L}}{dx}\right)^2
+ \left(\frac{d \phi_{\rm R}}{dx}\right)^2:.
\end{equation}
As expected, the energy density of the Gaussian model is rescaled by the function $f(x)$ under the SSD 
and the interaction strength is zero at $x=0$ and $x=\ell$. 
It should be stressed that some of the lattice models numerically studied in previous work, 
such as the spin-$\frac{1}{2}$ XXZ chain at criticality, fall into the category of $c=1$ CFT~\cite{Hikihara-Nishino} 
in the continuum limit. Therefore, given that the continuum limit of the lattice models with SSD 
is described by the Hamiltonian in Eq. (\ref{eq:gaussian_SSD}), we can explain why the SSD leaves 
the ground state (vacuum) of this class of systems almost unchanged. 
It remains open to investigate the effect of irrelevant interactions in lattice models on a (tiny) difference 
between the ground states of periodic and SSD Hamiltonians.

Let us now consider a more general case where the CFT has the affine Lie algebra 
as the spectrum generating algebra~\cite{YellowBook, Itzykson-Drouffe}. 
The Physical realization of this CFT is provided by the Wess-Zumino-Witten (WZW) model~\cite{Wess-Zumino, Witten}. 
There are also lattice models solvable by means of the Bethe ansatz and their 
effective field theories are described by the WZW model
~\cite{Affleck-Haldane, Affleck86, Affleck88, 
Kulish-Reshetikhin-Sklyanin, Takhtajan, Babudjan1, Babudjan2, Andrei-Johannesson, Johannesson, Sutherland}. 
The generators of the affine Lie algebra ${\hat {\frak g}}$ at level $k$ satisfy 
\begin{equation}
[J^a_n, J^b_m] = {\imi} f^{ab}_{c} J^c_{n+m} + k n \delta^{ab} \delta_{n+m,0},
\end{equation}
where $f^{ab}_c$ are the structure constants of the Lie algebra $\frak g$ 
and the level $k$ commutes with all the generators $J^a_m$. 
Note that the subalgebra $\{ J^a_0 \}_{a=1,2,...,{\rm dim}\frak g}$ constitute the Lie algebra $\frak g$. 
From the Sugawara construction of the energy-momentum tensor, 
the Virasoro generators can be expressed as
\begin{equation}
\label{eq:Virasoro_affine}
L_n = \frac{1}{2(k+h^{\lor})} \sum^{{\rm dim} \frak g}_{a=1} \sum_{m \in {\mathbb Z}}  : J^a_m J^a_{n-m} :, 
\end{equation}
where $h^{\lor}$ is the dual Coxeter number of 
%the Group $G$ associated to 
$\frak g$. 
The corresponding central charge is given by 
\begin{equation}
c = \frac{k\, {\rm dim}{\frak g}}{k+h^{\lor}}
\end{equation}
In the case of ${\frak g}=\frak{su}_N$, we have ${\rm dim}{\frak g}=N^2-1$ and $h^{\lor}=N$. 

We are now ready to show that the vacuum state $\ket{\rm vac}$ is an exact ground state of 
${\cal H}_{\rm SSD}$ when $L_n$ are defined as in Eq. (\ref{eq:Virasoro_affine}). 
Along the same lines as the $c=1$ CFT, we can rewrite ${\cal H}_{\rm L}$ in Eq. (\ref{eq:CFT_L_ham}) as
\begin{eqnarray}
{\cal H}_{\rm L} &=& \frac{\pi}{2 \ell (k+h^\lor)} \sum^{{\rm dim}\frak g}_{a=1} 
\sum_{n \ge 0} (J^a_{-n}-J^a_{-n-1}) (J^a_n-J^a_{n+1})
\nonumber \\
&=& \frac{\pi}{2 \ell (k+h^\lor)} \sum^{{\rm dim}\frak g}_{a=1} 
\sum_{n \ge 0} (J^a_{n}-J^a_{n+1})^\dagger (J^a_n-J^a_{n+1}), 
\end{eqnarray}
where we have used the fact that $(J^a_n)^\dagger=J^a_{-n}$. 
As is obvious, ${\cal H}_{\rm L}$ is written in terms of the sum of positive semidefinite operators 
and the same holds for ${\cal H}_{\rm R}$. 
Therefore, the ground state energy of the SSD Hamiltonian is again bounded below by $E_0/2$, 
which yields that $\ket{\rm vac}$ is a ground state of ${\cal H}_{\rm SSD}$.  
Again, it should be noted that our argument does not exclude the possibility that 
there are further zero-energy ground states other than $\ket{\rm vac}$. 
We finally remark that for a special class of CFTs with superconformal symmetry, 
the SSD Hamiltonian can be recast in the form $\{ Q, Q^\dagger \}$, 
where $Q$ and $Q^\dagger$ are nilpotent supercharges. 
A further discussion is given in Appendix B.

%%%%%%%%%%%%%%%%%%%%%%%%%%%%%%%%%%%%%%%%%%%%%%%%%%%%%%%%%%%%%%%%%
%%%%% conclusion   %%%%%
\section{Concluding remarks}
\label{sec:conclusion}

In conclusion, we have investigated solvable spin chains, the massless Dirac model, 
and conformal field theories with SSD. 
For the isotropic XY and the transverse field Ising chains at criticality, 
the unique ground state of an open chain with SSD is obtained as the ground state 
of a uniform chain with PBC, i.e., the SSD leaves the periodic ground state unchanged. 
%the ground state remains unchaged after the deformation. 
For the model of massless Dirac fermions, we showed that the Hamiltonian 
for the system with SSD is positive semidefinite, 
and found that the Dirac sea of the uniform system remains an exact ground state. 
For generic CFTs, we saw how the Hamiltonian of a system with SSD is 
expressed in terms of the $SL(2, \mathbb{C})$ subalgebra of the Virasoro algebra. 
This allowed us to show that the vacuum state, which is invariant under $SL(2, \mathbb{C})$, 
is an exact eigenstate of the SSD Hamiltonian. 
Furthermore, for a certain class of CFTs associated with affine Lie (Kac-Moody) algebras, 
it is revealed that the Hamiltonian is expressed as a positive semidefinite operator, 
similar to the case of the massless Dirac fermions. 
Thereby, we showed that the vacuum state is not only an eigenstate but also 
a ground state of the SSD Hamiltonian for this class of CFTs, including $c=1$ Gaussian model. 
Although we are not able to prove that there are no further ground states, 
our findings provide a reasonable explanation for the previous observations in numerical studies that the SSD 
well suppresses the effect of open boundaries on the ground state properties 
in a wide class of critical one-dimensional systems. 

It should be stressed that the key to understanding the remarkable correspondence 
observed in all the examples is that the chiral Hamiltonians, in terms of which 
the SSD Hamiltonian is expressed, annihilate the ground state of the uniform Hamiltonian, 
i.e., it is a zero-energy eigenstate of the chiral Hamiltonians. 
This strongly suggests that the original, chiral, and SSD Hamiltonians have 
further common eigenstates other than the ground state. 
In fact, we have some numerical evidence that the uniform and SSD XY chains 
have common excited eigenstates. This might be understood from 
the perspective of the theory of Macdonald polynomials, as suggested in previous work~\cite{Katsura_JPA}. 
It is also interesting to ask whether we can obtain some of eigenstates other than the vacuum state 
or even the whole spectrum of the SSD Hamiltonian for a given CFT. 
Recently, it has been shown that the simultaneous eigenstate of $L_0$ and $L_1$ of the Virasoro algebra 
(the coherent state of $L_1$) plays an important role in connection with the relationship between 
4-dimensional gauge theory and 2-dimensional CFT~\cite{AGT}.  
Therefore, it is natural to ask whether we can diagonalize linear combinations of $L_0$ and $L_{\pm 1}$, 
including the SSD Hamiltonian for generic CFTs. 
Finally, it should be mentioned that conformal mappings relating the SSD and the periodic Hamiltonians 
may not be restricted to the one used in the paper. 
Different mappings may provide a CFT interpretation of other deformations 
such as the hyperbolic~\cite{Ueda-Nishino08,Ueda-Gendiar10,Ueda-Nakano10}, 
exponential~\cite{Okunishi-Nishino10}, and sinusoidal~\cite{Gendiar2} deformations.

%%%%%%%%%%%%%%%%%%%%%%%%%%%%%%%%%%%%%%%%%%%%%%%%%%%%%%%%%%%%%%%%%
%%%%% acknowledgment   %%%%%
\ack{}
The author would like to thank 
Toshiya Hikihara, Isao Maruyama, Tomotoshi Nishino, and Shintarou Yanagida   
for their valuable comments and suggestions. 
The author was supported in part by Grant-in-Aid for Young Scientists (B) (23740298).

%%%%%%%%%%%%%%%%%%%%%%%%%%%%%%%%%%%%%%%%%%%%%%%%%%%%%%%%%%%%%%%%%
%%%%% Appendix   %%%%%
\appendix
\section{Lieb-Schultz-Mattis argument}
In this appendix, we extend the Lieb-Schultz-Mattis argument~\cite{Lieb-Schultz-Mattis} to 
the isotropic XY (or an equivalent free-fermion) chain with SSD 
and find that the lowest excitation energy is $O(1/L)$. 
As we will see below, the proof does not rely on the translational invariance of the Hamiltonian 
but on that of the ground state. 
Therefore, with a reasonable assumption that the ground state of the SSD Hamiltonian is 
almost exactly translationally invariant as observed in the previous numerical studies, 
it is plausible that the same argument holds for interacting cases such as the XXZ chain.  
This implies the gapless excitation in the limit of $L \to \infty$ when the SSD is applicable. 
%We study the nature of the excitation spectrum and prove ...

Let us consider the spin-$\frac{1}{2}$ isotropic XY chain in a field with SSD:
\begin{equation}
{\cal H}_{\rm SSD} = -J\sum^L_{j=1} f_{j+\frac{1}{2}} 
(S^x_j S^x_{j+1} + S^y_j S^y_{j+1}) -h\sum^L_{j=1} f_j S^z_j, 
\end{equation}
where $J>0$. 
%We prove the following theorem for the excitation spectrum.
%\\
%{\it Theorem}: 
Let $S^z_{\rm tot} := \sum^L_j S^z_j$. Since $S^z_{\rm tot}$ commutes with ${\cal H}_{\rm SSD}$, 
${\cal H}_{\rm SSD}$ is block-diagonal in the eigenvalue $M$ of $S^z_{\rm tot}$. 
It follows from the Perron-Frobenius theorem that the ground state in each $M$ sector is unique.  

We now prove that the lowest excitation energy is $O(1/L)$ in each $M$ sector. 
To show this, we follow \cite{Affleck-Lieb} and introduce the twist operator
\begin{equation}
U = e^{\imi {\cal A}},~~~{\cal A}=\frac{2\pi}{L} \sum^L_{j=1} j S^z_j
\end{equation}
and construct a trial state as $U \ket{\Psi_{\gs}}$, 
where $\ket{\Psi_{\gs}}$ denotes the ground state of the sector with fixed $M$.  
A straightforward calculation yields
\begin{eqnarray}
\fl~~~~~
U^\dagger {\cal H}_{\rm SSD} U -{\cal H}_{\rm SSD} &=& -J(\cos \delta -1 ) \sum^L_{j=1} f_{j+\frac{1}{2}} (S^x_j S^x_{j+1} + S^y_j S^y_{j+1}) 
+\imi \frac{\sin \delta}{\delta} [{\cal H}_{\rm SSD}, {\cal A}],
\nonumber \\
%\nonumber \\
%&+& \sin \delta \sum^L_{j=1} f_{j+\frac{1}{2}} (S^x_j S^y_{j+1} - S^y_j S^x_{j+1}), 
\end{eqnarray}
where $\delta = 2\pi/L$. Although the second term would appear to give $O(1)$ contribution, 
$\langle \Psi | [{\cal H}_{\rm SSD}, {\cal A}] \ket{\Psi}=0$ for any eigenstate of ${\cal H}_{\rm SSD}$ and hence
\begin{eqnarray}
\label{eq:exc_eng}
\fl~~~~~
\langle \Psi_{\gs} | U^\dagger {\cal H}_{\rm SSD} U -{\cal H}_{\rm SSD} \ket{\Psi_{\gs}} 
&=& -J (\cos \delta -1 ) \sum^L_{j=1} f_{j+\frac{1}{2}} \langle \Psi_{\gs}| S^x_j S^x_{j+1} + S^y_j S^y_{j+1} \ket{\Psi_{\gs}}
\nonumber \\
&\le & \frac{\pi^2 J}{L}  + O(1/L^2). 
\end{eqnarray}
This implies the existence of gapless excitations in the limit $L \to \infty$ if the trial state is orthogonal to $\ket{\Psi_{\gs}}$. 
It should be noted that our proof does not rely on any symmetry of ${\cal H}_{\rm SSD}$. 
We can, therefore, apply the above argument to a large class of deformed models in which 
the function $f_{j+\frac{1}{2}}$ is $O(1)$. 
The above argument also applies to the interacting cases with 
the $S^z_j S^z_{j+1}$ coupling since it commutes with ${\cal A}$. 

Let us next show the orthogonality $\langle \Psi_{\gs} | U\ket{\Psi_{\gs}}=0$ and thus that 
the estimate (\ref{eq:exc_eng}) gives an upper bound on the excitation energy. 
To this end, we make use of the translational invariance of the ground state $\ket{\Psi_{\gs}}$ 
that is a direct consequence of the result shown in Ref. \cite{Katsura_JPA} and Sec. \ref{sec:lattice} of the present paper. 
Let $T$ be the translation operator by one site. 
Then, from the translational invariance of $\ket{\Psi_{\gs}}$ and the fact that $\exp(2\pi \imi S^z_1)=-1$, 
we have 
\begin{eqnarray}
\fl 
\langle \Psi_{\gs} | U \ket{\Psi_{\gs}} &=& \langle \Psi_{\gs} |T U T^{-1} \ket{\Psi_{\gs}}
= \langle \Psi_{\gs}| U \exp (2\pi \imi S^z_1) 
\exp ( -\imi \delta S^z_{\rm tot}) \ket{\Psi_{\gs}}
\nonumber \\
&=& -\exp(-2\pi \imi M/L) \langle \Psi_{\gs} | U\ket{\Psi_{\gs}}. 
\end{eqnarray}
Therefore, the orthogonality $\langle \Psi_{\gs} | U\ket{\Psi_{\gs}}=0$ holds except for $M=\pm L/2$, i.e., 
the fully polarized cases. 

%Under this transformation, we find
%\begin{equation}
%(S^x_j, S^y_j, S^z_j) \to (-S^x_j, S^y_j, -S^z_j). 
%\end{equation}
%We also find that the Hamiltonian is invariant under the inversion, i.e., 
%${S}^\alpha_j \to {S}^\alpha_{L+1-j}$. Let $R$ be a combination of the above two symmetry operations. 
%\textbf{[H.K.: I now realize that the above argument doesn't work... 
%Tasaki's trick does not also work since it uses the reflection symmetry 
%about the site!!]}

\section{SCFT with SSD}
In this appendix, we show that the Hamiltonian with SSD can be expressed 
in terms of fermionic generators in superconformal field theories (SCFT). 

Let us first consider the ${\cal N}=1$ superconformal algebra. 
An example of the lattice model whose continuum limit falls into this class is the tricritical Ising model with 
the central charge $c=\frac{7}{10}$. 
The algebra of ${\cal N}=1$ SCFT contains fermionic generators besides the Virasoro ones
and is defined by the following commutation and anti-commutation relations~\cite{YellowBook, Mussardo}:
\begin{eqnarray}
\lbrack L_m, L_n \rbrack &=& (m-n) L_{m+n} + \frac{c}{12} (m^3-m) \delta_{m+n,0},
\label{eq:1SCFT_1} \\
\lbrack  L_m, G_r \rbrack &=& \left( \frac{m}{2} - r \right) G_{m+r},
\label{eq:2SCFT_2} \\
\{ G_r, G_s \} &=& 2 L_{r+s} + \frac{c}{3} \left( r^2 - \frac{1}{4} \right) \delta_{r+s,0},
\label{eq:1SCFT_3}
\end{eqnarray}
where $m,n \in \mathbb{Z}$ while 
$r,s$ are either half-integer indices (Neveu-Schwarz sector) 
or else are integer indices (the Ramond sector). 
$G_r$ have the meaning of the Fourier components of the field $G(z)$ 
with the conformal dimension $\frac{3}{2}$. 
From Eq. (\ref{eq:1SCFT_3}), one easily finds $\{ G_{\frac{1}{2}}, G_{-\frac{1}{2}} \} = 2L_0$
in the Neveu-Schwarz sector while $\{ G_0, G_0 \} = 2(G_0)^2 = 2 L_0 -\frac{c}{12}$ for the Ramond sector. 
Since $(G_0)^2$ is positive semi-definite and $G_r \ket{0} =0$ for $r \ge -\frac{1}{2}$, 
the vacuum $\ket{0}$ cannot be in the Ramond sector for unitary CFTs. 
Note that we required $(G_r)^\dagger = G_{-r}$. 
Let us first focus on the SSD Hamiltonian for the Neveu-Schwarz sector. 
Similarly, using Eq. (\ref{eq:1SCFT_3}), 
we find that the Hamiltonian ${\cal H}_{\rm L}$ defined in Eq. (\ref{eq:CFT_L_ham}) is expressed as
\begin{equation}
{\cal H}_{\rm L} = \frac{\pi}{4\ell} \{ G_{\frac{1}{2}} - G_{-\frac{1}{2}},  G_{-\frac{1}{2}} - G_{\frac{1}{2}} \},
~~~~~({\rm Neveu-Schwarz}). 
\end{equation}
This immediately implies that the eigenvalues of ${\cal H}_{\rm L}$ is nonnegative. 
We have similar expressions for the right-moving part of the Hamiltonian (${\cal H}_{\rm R}$). 
Therefore, the ground state energy of the SSD Hamiltonian is bounded below by $-\frac{\pi c}{12 \ell}$ 
and $\ket{0}$ is an exact ground state in the Neveu-Schwarz sector, 
which is consistent with what we found for the CFTs associated with affine Lie algebras in Sec. 4. 
For the Ramond sector, we have
\begin{equation}
{\cal H}_{\rm L} = \frac{\pi}{4\ell} \{ G_0-G_1, G_0 - G_{-1} \} - \frac{\pi c}{24 \ell},
~~~~~({\rm Ramond}).
\end{equation}

Let us next examine the ${\cal N}=2$ SCFT. 
It has been shown that the continuum field theory describing 
the supersymmetric quantum critical lattice models has this symmetry~\cite{Fendley_PRL_2003, Fendley_JPA_2003}.
A detailed comparison of the lattice model and ${\cal N}=2$ SCFT is presented in \cite{Huijse_2011}.  
The ${\cal N}=2$ superconformal algebra contains two fermionic fields $G^\pm (z)$ with the conformal dimension $\frac{3}{2}$ 
and a U(1) current $J(z)$ with the dimension $1$, and is defined by the following relations.
\begin{eqnarray}
\lbrack L_m, L_n \rbrack &=& (m-n) L_{m+n} + \frac{c}{12} (m^3-m) \delta_{m+n,0},
\label{eq:2SCFT_1} \\
\lbrack L_m, G^\pm_r \rbrack &=& \left( \frac{m}{2} -r \right) G^\pm_{m+r},
\label{eq:2SCFT_2} \\
\lbrack L_m, J_n \rbrack &=& -n J_{m+n},
\label{eq:2SCFT_3} \\
\lbrack J_m, G^\pm_r \rbrack &=& \pm G^\pm_{m+r},
\label{eq:2SCFT_4} \\
\{ G^\pm_r, G^\mp_s \} &=& 2 L_{r+s} \pm (r-s) J_{r+s} + \frac{c}{3} \left( r^2-\frac{1}{4} \right) \delta_{r+s,0},
\label{eq:2SCFT_5} \\
\{ G^+_r, G^+_s \} &=& \{ G^-_r, G^-_s \} =0,
\label{eq:2SCFT_6} \\
\lbrack J_m, J_n \rbrack &=& \frac{c}{3} m \delta_{m+n,0},
\label{eq:2SCFT_7}
\end{eqnarray}
where $m,n \in \mathbb{Z}$. 
Similar to the ${\cal N}=1$ SCFT, $r,s \in \mathbb{Z}+\frac{1}{2}$ in the Neveu-Schwarz sector 
and $r,s \in \mathbb{Z}$ in the Ramond sector. 
Note that hermitian conjugation of $G^\pm_r$ is defined by $(G^\pm_{r})^\dagger = G^\mp_{-r}$. 
From the fact that the operators $G_r = (G^+_r+G^-_r)/\sqrt{2}$ generate the algebra 
isomorphic to ${\cal N}=1$ superconformal algebra, we find that 
the SSD Hamiltonian in the Neveu-Schwarz sector is expressed as
\begin{eqnarray}
{\cal H}_{\rm L} &=& \frac{\pi}{4 \ell} \{ Q^\dagger -Q, Q-Q^\dagger \} 
\nonumber \\
&=& \frac{\pi}{2 \ell} \{ Q, Q^\dagger \},
\label{eq:SUSYQM}
\end{eqnarray}
where $Q^\dagger=(G^+_{\frac{1}{2}}-G^+_{-\frac{1}{2}})/\sqrt{2}$ is the supercharge 
satisfying $(Q^\dagger)^2=Q^2=0$. 
One can explicitly confirm the above relation using Eq. (\ref{eq:2SCFT_6}). 
It is now apparent that ${\cal H}_{\rm L}$ in the Neveu-Schwarz sector is written 
as the anti-commutator of the supercharges as usual in the supersymmetric quantum mechanics~\cite{Witten_NPB}. 
Nice corollaries of this result are as follows: 
i) all the eigenvalues of ${\cal H}_{\rm L}$ are nonnegative, 
ii) States with a positive energy come in pairs, 
iii) any state with zero energy is a ground state and is annihilated by both $Q$ and $Q^\dagger$. 
It is interesting to note that in the original uniform theory (${\cal H}_0$), this supersymmetric structure is realized not in 
the Neveu-Schwarz sector but in the Ramond sector \cite{Huijse_2011}. 
Let us finally consider the SSD Hamiltonian for the Ramond sector. 
Again using the isomorphism, we have
\begin{eqnarray}
{\cal H}_{\rm L} &=& \frac{\pi}{8 \ell} \{ G^+_0 + G^-_0 - G^+_1-G^-_1, G^-_0 + G^+_0 - G^-_{-1}-G^+_{-1} \} 
- \frac{\pi c}{24 \ell}
\nonumber \\
 &=& \frac{\pi}{8 \ell} \{ G^+_0 - G^+_1, G^-_0 - G^-_{-1} \} + \{ G^+_0 - G^+_{-1}, G^-_0 - G^-_1 \} - \frac{\pi c}{24 \ell}.
\end{eqnarray}
In this case, ${\cal H}_{\rm L}$ cannot be written in a simple form like Eq. (\ref{eq:SUSYQM}).

%%%%%%%%%%%%%%%%%%%%%%%%%%%%%%%%%%%%%%%%%%%%%%%%%%%%%%%%%%%%%%%%%
\section*{References}
\providecommand{\newblock}{}

\end{document}